\documentclass[a4paper,12pt]{article}
\date{}
\author{}
\title{\LARGE\bf Minimal vertex covers of random trees}

\usepackage{t1enc}
\usepackage[latin1]{inputenc}
\usepackage[final]{graphicx}
\usepackage{amsfonts}

\begin{document}
\maketitle

\vspace{-1.2cm}

\centerline{\large St\'ephane Coulomb\footnote[1]{Email:
    coulomb@spht.saclay.cea.fr}}

\vspace{.3cm}

\centerline{\large Service de Physique Th\'eorique de
  Saclay\footnote[2]{\textit{ Laboratoire de la Direction des Sciences
      de la Mati\`ere du Commisariat \`a l'Energie Atomique, URA2306
      du CNRS}}}

\vspace{.3cm}

\centerline{\large CE Saclay, 91191 Gif sur Yvette, France}

\vspace{.3cm}

\begin{abstract} 
We study minimal vertex covers of trees. Contrarily to the number
$N_{vc}(A)$ of minimal vertex covers of the tree $A$, $\log N_{vc}(A)$
is a self-averaging quantity. We show that, for large sizes $n$,
$\lim_{n\to +\infty} <\log
N_{vc}(A)>_n/n= 0.1033252\pm 10^{-7}$. The basic idea is, given a
tree, to concentrate on its degenerate vertices, that is those
vertices which belong to some minimal vertex cover but not to all of
them. Deletion of the other vertices induces a forest of totally
degenerate trees. We show that the  problem reduces to the computation
of the size distribution of this forest, which we perform
analytically, and of the average $<\log N_{vc}>$ over totally
degenerate trees of given size, which we perform numerically.
\end{abstract}

\section{Introduction}
The vertex-cover problem, as other combinatorial problems, is arousing 
growing interest in the fields of statistical physics and disordered
systems. In particular, it helps to understand, and the machinery of
optimization algorithms helps to solve, spin-glasses and random
hamiltonian models (see \cite{hw} for a recent review of the problem,
\cite{bg} for a critical analysis point of view). A possible question is : given a graph, what can
be said about the size and number of its minimum vertex covers ?
Another approach consists in answering this question \textit{on
  average}, for a given statistical ensemble of graphs.

In this paper, we are concerned with average behaviour, and focus on
the simple situation of trees. In this case, good algorithms are
known, for instance based on the so-called b-colorings (see
\cite{coubau} and sect.\ref{prelim}), to find the
number $N_{vc}(A)$ or size of minimal vertex covers of a given tree
$A$. In fact, if each tree of given size $n$ has the same probability,
then the average number of vertex covers can also be retrieved analytically 
by means of these b-colorings. However, this is not a self-averaging quantity for
large $n$, and it would be desirable to find a thermodynamically
extensive quantity giving a somewhat more physical insight into the
number of minimal configurations of a random tree.

We claim that $<\log N_{vc}(A)>_n$ is indeed self-averaging, and the
reason is as follows. Suppose that we delete from $A$ all the vertices 
which are not degenerate (that is, those which belong either to all
the minimal vertex covers of $A$ or to none of them). Then we obtain (see
sect.\ref{redforest}) a forest with the same number of minimal vertex
covers as $A$ and whose vertices are all degenerate. Moreover, in this 
forest, the number of trees of given size scales thermodynamically
with the size of $A$ (see sect.\ref{sizedistr}), and the probability
of appearance of a given tree depends only on its size. In other
words, as far as we are concerned with the number of minimal vertex
covers, picking at random a tree on $n \gg 1$ vertices amounts for
each $i\geq 1$  to picking with uniform law  $c_i n$ totally
degenerate trees on $i$ vertices. And, in turn, it is expected that
such a typical tree $A$ verifies $\log N_{vc}(A)\approx n \sum_i c_i <\log
N_{vc}>_i^R$, where  $<\log N_{vc}>_i^R$ is the average of $\log
N_{vc}$ over totally degenerate trees on $i$ vertices.

The computation thus reduces to that of the scaling parameters $c_i$
for the size distribution, and to the evaluation of the average of
$\log N_{vc}$ over totally degenerate trees with given size (see
sect.\ref{covred}). 

But let us begin with a remainder of some basic facts and the crucial
theorem on b-colorings.

\section{Preliminary observations}
\label{prelim}
\subsection{Basic definitions}
A \textit{graph} is a pair $A=(V,\mathcal E)$ where
$V$ is a set with $n\geq 1$ elements (written $|V|=n$ in the sequel)
and $\mathcal E$ is a subset of $\{\{x,y\}\subset V;x\neq y\}$. $V$ is
the set of \textit{vertices} of $A$ and $\mathcal E$ the set of
\textit{edges} of $A$, $n$ is the \textit{size} of $A$, denoted by
$|A|$. In this paper $A$ is called a \textit{labeled graph} if $V$
consists of positive integers.

Given two distinct vertices $x,y$ of the graph $A=(V,\mathcal E)$, a
path from $x$ to $y$ in $A$ is a sequence
$\{v_0,v_1\},\{v_1,v_2\},\cdots,\{v_{p-1},v_p\}$ of edges of $A$ such
that $v_0=x,v_p=y$ and $v_i\neq v_j$ if $i\neq j$. A graph is called a
\textit{tree} if any two distinct vertices are connected by a unique
path, and a \textit{forest} if any two distinct vertices are connected
by at most one path.

A \textit{rooted tree} is a triple $(V,\mathcal E,r)$, such that $(V,\mathcal E)$ is a tree and $r\in V$.

A \textit{vertex cover} of the graph $A=(V,\mathcal E)$ is a subset of
$V$ containing at least one end of each edge of $A$. A vertex cover of 
$A$ is \textit{minimal} if there does not exist any other vertex cover
with less elements. In the sequel, the number of minimal vertex covers 
of $A$ is denoted $N_{vc}(A)$.

\subsection{Some useful results}

\paragraph{The exponential generating function of rooted trees} It is
defined as $T(x)=\sum_A \frac{x^{|A|}}{|A|!}$, where the sum runs over
all rooted trees. Cayley's formula states that the number of rooted
trees on $n$ vertices is $n^{n-1}$, hence $T(x)=\sum_{n\geq
  1}\frac{n^{n-1}}{n!}x^n$ and this implies that $T(x)=xe^{T(x)}$ as
can be deduced by a direct combinatorial argument relying on the recursive nature of
rooted trees.

\paragraph{A theorem on minimal vertex covers of trees} It has been
shown in \cite{coubau} that, for any tree $A=(V,\mathcal E)$, there 
exists a unique triple $(B,{\mathcal R},G) \subset V \times {\mathcal
  E} \times V$, called the \textit{b-coloring} of $A$, such that
\begin{itemize}
\item $B$,$G$ and the set of end-vertices of ${\mathcal R}$ form a
  partition of $V$.
\item The edges in ${\mathcal R}$ are non-adjacent; the edges
with one end-vertex in $G$ have the other end-vertex in $B$; each
vertex in $B$ is connected to $G$ by at least two edges.
\end{itemize}

Moreover, the b-coloring of $A$ has the following connection with its
minimal  vertex covers~: $B$ (resp. $G$) is the set of vertices
contained in all (resp. none) of the minimal vertex covers of
$A$. Consequently,  any end-vertex of $\mathcal R$ is contained in
some minimal vertex cover of $A$ but not in all of them~: these
vertices are called \textit{degenerate}.

An additional result is that any minimal vertex cover of $A$ contains
exactly one end-vertex of each edge in $\mathcal R$. Consequently, a
vertex cover of $A$ is minimal if and only if it contains
$|B|+|{\mathcal R}|$ vertices. 

In the sequel, vertices in $B$ and $G$ and end-vertices of $\mathcal
R$ will be called respectively \textit{brown}, \textit{green} and
\textit{red} vertices, while edges in $\mathcal R$ will be called red
edges. A tree with no brown or green vertices is said to be red.

\section{Red forest of a tree}
\label{redforest}
Given a tree $A=(V,\mathcal E)$ and a non-empty set $S\subset V$ of
vertices, the forest induced by $A$ on $S$ is defined as $(S,\mathcal
E')$, where $\mathcal E'$ consists of those edges in $\mathcal E$ with
both ends in $S$. 
If $A$ is a tree with b-coloring $(B,{\mathcal R},G)$, and such that
${\mathcal R}\neq \emptyset$, define the \textit{red forest} of $A$ to 
be the forest induced by $A$ on the set of red vertices. Denote
$A_1=(V_1,{\mathcal E}_1),\cdots,A_p=(V_p,{\mathcal E}_p)$ the trees
of that forest. Then it follows at once from the definitions that
$A_i$ has b-coloring $(\emptyset,{\mathcal R}\cap {\mathcal
  E}_i,\emptyset)$, hence is red. But if $C$ is a minimal vertex cover
of $A$, $C\cap V_i$ is a vertex cover of $A_i$. Since $C$ contains
exactly one end of each red edge of $A$, $C\cap V_i$ contains
$|{\mathcal R}\cap {\mathcal
  E}_i|$ vertices of $A_i$~: it is a minimal vertex cover of
$A_i$. Now, given minimal vertex covers $C_1,\cdots,C_p$ of the
$A_i$'s, $B\cup C_1\cup\cdots \cup C_p$ is a vertex cover of $A$
(because an edge of $A$ either is an edge of some $A_i$ or has at
least one end in $B$), which is minimal since it contains
$|B|+|{\mathcal R}|$ vertices. It is in fact the only minimal vertex
cover of $A$ which coincides with $C_i$ on each $A_i$, and this proves
that~: 
$$N_{vc}(A)=\prod_{i=1}^p N_{vc}(A_i).$$

Let us define the \textit{size distribution} of a forest $F$ as the 
sequence $D=(D_i)_{i\geq 1}$, where $D_i$ is the number of components 
of size $i$ in $F$. Given two forests $F_1,F_2$ of red trees, with
same size distribution $D$,
there is no difficulty in proving that the numbers of trees with red
forests respectively $F_1$ and $F_2$ are equal. In other words the
number of trees on $n$ vertices with given red forest $F$ depends on
$F$ only via its size distribution $D$~: this number shall
be denoted $\nu_D(n)$ in the sequel. Note that $\nu_D(n)=0$ if
$D_i\neq 0$ for some $i>n$.

If we denote by $\lambda_i$ the
sum over red trees $R$ on $i$ vertices of $\log N_{vc}(R)$, the
preceding remarks allow to write our sum over trees of size $n$ as
$$\sum_A \log N_{vc}(A)=\sum_D \nu_D(n) (D_1 \lambda_1+D_2
\lambda_2+\cdots+D_n \lambda_n).$$
We are thus led to the computation of the $\nu_D$'s and
$\lambda_i$'s. Note already that a red tree has even size,
whence $\nu_D(n)=0$ if $D_{2i+1}\neq 0$ for some $i$. We now come to
the analytic computation of $\nu_D(n)$.

\section{Size distribution}
\label{sizedistr}
Denote by $G,B,R$ respectively the exponential generating functions
for the number of rooted trees with root of color green, brown 
and red. For instance, $G(x)\equiv \sum_A \frac{1}{|A|!} x^{|A|}$, where the 
sum runs over all rooted trees with green root.
The following relations hold between these generating functions (see
\cite{coubau} for details and the combinatorial meaning of $U,Q$)
\begin{eqnarray*}
G & = & xe^U \\
U & = & x e^{B+R} (e^G-1) \\
B & = & x e^{B+R} (e^G-1-G) \\
R & = & xQe^{B+R} \\
Q & = & xe^{B+R},
\end{eqnarray*}
leading in particular to $B(x)=T(x)+T(-T(x))-T(-T(x))^2$. Now, let us 
look more closely at those trees with red root. The red forest of such 
a tree $A$ has exactly one component containing the root, and the size 
$s(A)$ of this component can be encoded in the following generating
function, where the sum runs over rooted trees with red root
$$R_0(x,y)\equiv \sum_A \frac{1}{|A|!} x^{|A|} y^{s(A)}.$$
Since $R_0=xyQ_0e^{B+R_0}$ with $Q_0=xye^{B+R_0}$, it follows that
$R_0=T(2x^2y^2e^{2B(x)})/2$, and the total number of red components of 
size $2p$ among labeled trees of size $n$ is
$$\frac{1}{2p}n![x]_n[y]_{2p}R_0(x,y)=\frac{(2p)^{p-2}}{p!}n![x]_{n-2p}e^{2pB(x)}$$ 

A straightforward application of the saddle-point method then shows
that, for large $n$, the average number of red components of size $2p$
scales thermodynamically with $n$~: $C_{2p}(n) \sim c_{2p}n$ and
\begin{equation}
c_{2p}=\frac{(2p)^{p-1}}{p!}T'T^{2p-1}e^{-2pT^2}(2T^2-1),
\label{c2p}
\end{equation}
where, in the above formula, $T(x)$ and its derivative are taken at the
saddle-point $x=-1$.
For large $p$, we get that $\frac{\log c_{2p}}{p}$ tends to $\log (2eT^2\exp(-2T^2))
\approx -0.0844424236$, showing that $c_{2p}$ decays exponentially.

Now, we make the ``thermodynamic limit'' assumption that the number of 
trees with given size in the red forest of some random large tree is a 
self-averaging quantity. That is, we suppose that, for large $n$, the
trees which contribute significantly to $< \log N_{vc}>_n$ have indeed
$C_2$ trees of size 2, $C_4$ trees of size 4,$\cdots$. The
distribution $\nu_D$ hence becomes irrelevant, since it concentrates
on one particular value, and the average becomes
$$\lim_{n\to +\infty} <\log N_{vc}>_n/n=\sum_i c_{2i} \frac{\lambda_{2i}}{N_{2i}},$$
where $N_{2i}$ denotes the number of red trees of size $2i$.

\section{Minimal vertex covers of the red trees}
\label{covred}
In this section, we compute analytically the number $N_{2p}$ of red trees on
$2p$ vertices and give a numeric estimate of the $\lambda_{2i}$'s. In
fact, one could deduce directly $R(x)$, whence the $N_{2p}$'s, from
the set of equations on $B,G,R,Q,U$ stated in the preceding
section. But we prefer to give a direct combinatorial derivation
which, after slight adaptations, shall give also the total number of
minimal vertex covers among red trees of size $2p$.

\subsection{Overview}
As was already emphasized, a red tree $A$ has an even number of vertices, say
$2p$, and we associate to $A$ its \textit{shrinked
  tree} $\tilde A$ as follows
\begin{itemize}
\item The vertices of $\tilde A$ are the red edges of $A$, so $\tilde
  A$ has size $p$ ;
\item Two vertices of $\tilde A$ are connected in $\tilde A$ if and
  only if the corresponding two red edges of $A$ are connected by some 
  other edge in $A$.
\end{itemize}
This procedure is uniquely defined and, if the set of vertices of $A$
is $V$, that of its shrinked tree is a partition of $V$ into sets of 2
elements. Such a partition will be called a \textit{pairing} of $V$~:
note that it consists of the red edges of $A$.

Conversely, let $V$ be a set ($|V|=2p$). There are
$\frac{(2p)!}{2^pp!}$ pairings of $V$, and $p^{p-2}$ trees with set of 
vertices equal to one of these pairings. Given such a tree $B$, the
number of red trees on $V$ with shrinked tree $B$ is 
$4^{p-1}$, because each of the $p-1$ edges of $B$ leaves 4
possibilities for the corresponding edge of the red tree. Hence, the
number of red trees on $2p$ vertices is
$$N_{2p}=\frac{(2p)!}{p!}(2p)^{p-2},$$
so the number $2pN_{2p}$ of rooted trees has exponential
generating function $R(x)=T(2x^2)/2$.

Let us now enumerate the total number of minimal vertex covers among 
the red trees of size $2p$. Consider a minimal vertex cover on a
labeled tree $A$ of size $2p$. To encode this vertex cover, add an
arrow at each covered end of each black edge (that is, each edge which 
is not red). By definition of vertex covers
a black edge is either oriented (one arrow) or bi-oriented (two arrows). 

Now, we apply the shrinking procedure as defined above, but we keep
track of the orientations~: this leads to a tree on $p$ vertices, each
edge being either oriented or bi-oriented.

Again this procedure is uniquely defined. If $V$ is a set on $2p$
vertices, the number of trees with set of vertices a pairing of $V$
and with edges either oriented or bi-oriented is
$\frac{(2p)!}{2^pp!}p^{p-2}3^{p-1}$. Given one such tree $B$, the
number of covered red trees $A$ with shrinked tree $B$ is $2^p$. Indeed,
each of the $p$ vertices of $B$ corresponds to a red edge of $A$,
which may be covered in two ways. Once this choice has been made,
the way the black edges connect the red edges with each other is
completely constrained by their (bi-)orientation.

Hence, the total number of minimal
vertex covers over red trees of size $2p$ is
$3\frac{(2p)!}{p!}(3p)^{p-2}$, and the average number of minimal
vertex covers among red trees on $2p$ vertices is
$<N_{vc}>_{2p}^R=2(3/2)^{p-1}$. 
\subsection{Theoretical viewpoint}

Both for theoretical understanding and for numerical purpose, it
proves useful to focus on rooted trees, and we denote by
$n_+(A)$ (resp. $n_-(A)$) the number of minimal vertex covers which
contain (resp. do not contain) the root of the rooted red tree $A$.

A red tree with root $r$ may be seen recursively as an edge
$\{r,r'\}$, with both ends connected to the root of arbitrarily many
red rooted trees. And it is clear (see \cite{coubau} for details) that
a set $S$ of vertices of $A$ is a minimal vertex cover of $A$ if and
only if~: (i) it induces a minimal vertex cover on each of these
attached subtrees (ii) exactly one end of $\{r,r'\}$ is not in $S$
(iii) the edges incident at this vertex have the other end in $S$.
Consequently, denoting by $A_i$ the red trees attached to $r$ and by
$A'_j$ those attached to $r'$~: 
\begin{eqnarray}
\label{rec1}
n_+(A) & = & \prod (n_+(A_i)+n_-(A_i)) \prod n_+(A'_j) \\
\label{rec2}
n_-(A) & = & \prod n_+(A_i)\prod (n_+(A'_j)+n_-(A'_j))
\end{eqnarray}

Now, let us have a closer look at the generating function for rooted
red trees $R(x)=T(2x^2)/2$. As follows from the equation for $T$, $R$
should be such that $R(x)=x^2e^{2R(x)}$. Combinatorially, this means that
the number of rooted red trees on $2p$ vertices is
\begin{equation}
(2p)![x^{2p-2}]\left(\sum_{k\geq 0} \frac{1}{k!}\left[\sum_A
\frac{x^{|A|}}{|A|!}\right]^k\right)\left(\sum_{k'\geq 0} \frac{1}{k'!}\left[\sum_A
\frac{x^{|A|}}{|A|!}\right]^{k'}\right),
\label{nbred}
\end{equation}
 where $A$ ranges over rooted
red trees. But building a rooted tree on $n$ vertices amounts to
choosing (i) the root $r$ and the vertex $r'$ with whom $r$ shares its
red edges ($2p(2p-1)$ ways) (ii) the numbers $k$ and $k'$ of rooted
trees attached respectively to those vertices (iii) those trees themselves
$A_1,\cdots,A_k$ and $A'_1,\cdots,A'_{k'}$, in such a way that their
total number of vertices is $2p-2$ (iv) finally, a relabeling of those
trees which exhausts the labels $\neq r,r'$ ($(2p-2)!/(\prod
|A_i|!\prod |A'_j|!)$ ways). Each term of the expansion of $\left[\sum_A
\frac{x^{|A|}}{|A|!}\right]^k$ corresponds to a particular
\textit{ordered} choice in (iii), and the $1/k!$ factor just gets rid
of this ordering. This is true also for the primed term, hence the
combinatorial meaning of the equation for $R$ is clear and we now
apply it to our vertex covers problem.

The set $S$ of functions $\mathbb{N}^2\rightarrow \mathbb{R}$ is a vector
space. If $\phi$ is such a function, and
$\phi(a,b)=x_{ab},a,b\in\mathbb{N}$, we write
$\phi=\sum_{a,b}x_{ab}(a,b)$. If $\psi=\sum_{a,b}x'_{ab}(a,b)$ is
another function, let their product be
$\phi*\psi=\sum_{a,b,a',b'}x_{ab}x'_{a'b'}(aa',bb')$. $S$ is then an
algebra, generated by the $(a,b),a,b\in\mathbb{N}$. Let $\sigma$ be
the (algebra) morphism such that $\sigma(a,b)=(b,a)$ for all $a,b$ and $\rho$
the (vector space) morphism such that $\rho(a,b)=(a+b,a)$. Then
eqs.(\ref{rec1},\ref{rec2}) rewrite $(n_+(A),n_-(A))=\prod \rho
(n_+(A_i),n_-(A_i)) * \prod \sigma \rho (n_+(A'_j),n_-(A'_j))$. Hence,
our remarks on the combinatorial meaning of eq.(\ref{nbred}) show that
the formal power series $R_{+-}(x)\equiv\sum_A
(n_+(A),n_-(A))\frac{x^{|A|}}{|A|!}$ obeys the equation
$$R_{+-}(x)=x^2 e^{\rho R_{+-}(x)+\sigma\rho R_{+-}(x)}.$$ Of course,
in this equation, the exponential is defined by its power series, the
product being as defined above.

Let $f_{lm}$ be the (algebra) morphism such that $f_{lm}(a,b)=a^lb^m$ for
all $a,b$. Then $f R_{+-}(x)=x^2 e^{f\rho R_{+-}(x)+f\sigma\rho
R_{+-}(x)}$, so we have the following generating functions for rooted trees:
\begin{eqnarray}
R_{lm}(x) & \equiv & \sum_A n_+(A)^l n_-(A)^m \frac{x^{|A|}}{|A|!}
\nonumber \\
 & = & x^2 \exp \left( \sum_{k=0}^l {l \choose k}
   R_{k,l+m-k}+\sum_{k=0}^{m} {m \choose k} R_{k,l+m-k}
 \right)
\label{rec}
\end{eqnarray}

For the first two values of $n$, the resulting system of equations is
easily solved. For instance~:
\paragraph{For $l=m=0$~:} $R_{0,0}(x)=x^2 e^{2R_{0,0}(x)}$, so
  $R_{0,0}(x)=T(2x^2)/2$ as expected.
\paragraph{For $l=1,m=0$ or $l=0,m=1$~:} $R_{1,0}(x)=R_{1,0}(x)=x^2 e^{3R_{1,0}(x)}$, so
  $R_{1,0}(x)=T(3x^2)/3$, again in agreement with the formula
  above.

And this seems to be the largest value of $n$ for which the exact
solution functions are retrievable. In the case where $l+m=2$, the
system reduces to an implicit expression for $R_{2,0}$~: $R_{2,0}=x^2\exp
\left(3R_{2,0}+2R_{2,0}e^{-R_{2,0}} \right)$, still allowing
asymptotic computations. However, we have not found a systematic
treatment for the study of eq.(\ref{rec}) which would have been a
possible starting point for the replica method.

\begin{table}
  \centering \tabcolsep 15pt
  \begin{tabular}{c|ccccc}
    $k$ & 1 & 2 & 3  \\ \hline
    $M_k$  & 0.20273  & 0.41576 &  0.63658 \\
  \end{tabular}
  \caption{\em Moments $M_k=\lim_{p\to +\infty} \frac{1}{2p}\log <N_{vc}^k>_{2p}^R$
of the number of minimal vertex covers of red trees, as obtained from eq.(\ref{rec}).}
  \label{mom}
\end{table}

We now come to the numerical evaluation of the $\lambda_{2p}$'s.

\subsection{Numerical computations}
Given a red tree $A$ on $2p$ vertices, one can choose any of its vertices
as a root and apply recursively equations (\ref{rec1},\ref{rec2}) to
compute $n_+(A),n_-(A)$ in $O(p)$ time. However, the number of such
trees increases exponentially with $p$, and systematic enumeration
soon becomes a challenge.

For small trees ($p\leq 16$), we compute the
exact distribution of the number of  minimal vertex covers. The
algorithm is based on an exhaustive recursive enumeration of rooted
trees \cite{lr}, followed by systematic unshrinking.

For larger trees, we proceed as follows. The number of red trees with
given shrinked tree $A$ depends only on $|A|$, and every red tree on
$2p$ vertices has a unique shrinked tree,  which is of size
$p$. Hence, to pick randomly a red tree on $2p$ vertices with uniform
law, it suffices to~: (i) Pick randomly a tree $A$ on $p$ vertices,
with uniform law (this is conveniently done by means of the Prüfer
bijection between those trees and sequences of
$\{1,\cdots,p\}^{\{1,\cdots,p-2\}}$) (ii) Choose, again with uniform
probability, one of the red trees with shrinked tree $A$.

The number of samples picked for each size was chosen so as to ensure
a precision of $10^{-7}$ on $<\log N_{vc}>/n$. From the fact that
$<\log N_{vc}>_n/n=\sum_i c_{2i}\lambda_{2i}/N_{2i}$, it follows that
an error $\delta_{2i}$ on $\frac{1}{2i}\frac{\lambda_{2i}}{N_{2i}}$
leads to a maximum error $\sum_i c_{2i}2i\delta_{2i}$ on $<\log
N_{vc}>_n/n$. From eq.(\ref{c2p}) we see that $c_{2i}$ decays
exponentially fast with $i$~: in practice, we took $8\ 10^9$ samples
for each size $17\leq p \leq 45$ and $1.5\ 10^8$ samples for sizes $46\leq
p \leq 189$. And this leads to $$\lim_{n\to +\infty} <\log
N_{vc}(A)>_n/n =\sum_{p>0} c_{2p}<\log N_{vc}>_{2p}^R=0.1033252\pm10^{-7}$$

Those numerical simulations also give evidence that, for red trees of
large size $2p$, the random variable $X_p=(\log N_{vc})/(2p)$ is
self-averaging. Indeed, for each of the sizes considered in the previous
paragraph, it is possible to get the approximate distribution of
$X_p$, and it appears that $(X_p-<X_p>)\sqrt{p}$ approaches a fixed
gaussian distribution for large $p$. Numerically, we find $\lim_{p\to
+\infty} <X_p>=\lim_{p\to  +\infty} <\log N_{vc}>_{2p}^R/2p=0.1963\pm
10^{-4}$, to be compared with the first few moments of table
[\ref{mom}]. In fact, approximating the first few $M_k$'s by a rational function leads 
to estimate $\lim_{p \to \infty} <X_p>=\frac{dM_k}{dk}|_{k=0}\approx
0.196$, a result remarkably close to the expected limit. Good
understanding of this self-averaging feature would certainly be a
crucial issue in the exact derivation of $\lim_{p\to +\infty} <X_p>$,
and presumably also of the corresponding limit for general trees.
\\
\\
\textit{I am very grateful to  Michel Bauer for interesting remarks
  and discussions.}


\begin{thebibliography}{}
  
\bibitem{hw} A. K. Hartmann, M. Weigt, \textit{Statistical mechanics of 
    the vertex-cover problem}, J. Phys. A: Math. Gen. \textbf{36}
  (2003) 11069-11093.
\bibitem{bg} M. Bauer, O. Golinelli, \textit{Core percolation in
    random graphs~: a critical phenomena analysis}, Eur. Phys. J. B
  \textbf{24} (2001), 339-352 , \texttt{cond-mat/0102011}.
\bibitem{coubau} S. Coulomb, M. Bauer, \textit{On vertex
    covers, matchings and random trees}, \texttt{cond-mat/0407456}
\bibitem{lr} G. Li, F. Ruskey, \textit{The Advantages of Forward
    Thinking in Generating Rooted and Free Trees}, SODA 1999: 939-940.
\end{thebibliography}
\end{document}